\documentclass[10pt,onecolumn]{article}

\usepackage{graphicx}
\usepackage{dcolumn}
\usepackage{bm}
\usepackage{hyperref}
\usepackage{tikz}
\usepackage{subfig}
\usepackage{amsmath}
\usepackage{amssymb}
\usepackage{authblk}
\usepackage{geometry}
\DeclareMathOperator{\im}{\rm{Im}}

\title{Limitations of Clustering Using Quantum Persistent Homology}

\author[1]{Niels M. P. Neumann}
\author[1]{Sterre R. den Breeijen}
\affil[1]{%
 Department of Cyber Security and Robustness,
 The Netherlands Organisation for Applied Scientific Research,
 The Hague, The Netherlands. 
 \textit{E-mail address}: \{niels.neumann, sterre.denbreeijen\}@tno.nl
}%
\date{}                     
\providecommand{\keywords}[1]{\textbf{\textit{Index terms---}} #1}

\begin{document}
\maketitle
\begin{abstract}
Different algorithms can be used for clustering purposes with data sets. On of these algorithms, uses topological features extracted from the data set to base the clusters on. The complexity of this algorithm is however exponential in the number of data points. Recently a quantum algorithm was proposed by Lloyd Garnerone and Zanardi with claimed polynomial complexity, hence an exponential improved over classical algorithms. However, we show that this algorithm in general cannot be used to compute these topological features in any dimension but the zeroth. We also give pointers on how to still use the algorithm for clustering purposes. 
\end{abstract}

\section{\label{sec:introduction}Introduction}
The amount of data annually processed worldwide is ever increasing. Machine learning uses this data to perform specific tasks, such as speech recognition \cite{Hinton_etAll_12} and search engines \cite{Joachims02}. However, with the increasing amount of data, also the complexity of these algorithms increases and they quickly become intractable. It is here where the relatively new field of quantum computing might provide a solution. 

Apart from the breakthrough algorithms of Shor \cite{Shor94} and Grover \cite{Grover96}, machine learning is one of the fields where quantum computing might excel at. However, while theoretical promising, most quantum machine learning algorithms have often overlooked caveats, as indicated by \cite{Aaronson15}. The same work however also points out a promising example of a quantum machine learning algorithm: using a topological algorithm for data analysis \cite{LloydGZ16}.

This algorithm uses persistent homology, a concept first introduced in \cite{ZomorodianC04}. Clustering algorithms using this persistent homology have later been proposed by \cite{IslambekovG18} and \cite{PereiraM15}, as well as \cite{WubieAGHMKH18}, a surveying paper that presents two other such algorithms. 

These algorithm are however classical and face computational problems. Namely, computing the peristent homology requires constructing simpilicial complexes, the size of which may be exponential in the number of data points. Especially for this reason, a quantum algorithm might bring the solution, as it might give an exponential speedup, resulting in a polynomial-time algorithm. 

The algorithm given in \cite{LloydGZ16} indeed claims an exponential speedup over classical algorithms. However, the output of the proposed algorithm is the homology of the data set, after which it is concluded that the persistent homology can be extracted from it. We show however that in general it is not possible to compute the persistent homology using this algorithm in dimensions other than the zeroth. The results given in this paper as well as a more formal introduction to simplicial homology is given in \cite{MLCBN19}.

In Sec.~\ref{sec:computingPH} persistent homology is explained and how to compute it in general. Sec.~\ref{sec:quantumAlgorithmPH} briefly explains the algorithm of \cite{LloydGZ16} and Sec.~\ref{sec:LimitationsOfAlgorithm} shows that in general the algorithm cannot be used to compute the persistent homology for dimensions higher than zero, while it also presents pointers on how to still use the algorithm for clustering tasks. Sec.~\ref{sec:conclusion} gives a summary and final remarks. 

\section{\label{sec:computingPH}Computing the persistent homology}
We briefly introduce the concept of simplicial homology, then, using point clouds, we will introduce the concept of persistent homology without formal definitions. For more details we refer to \cite{ZomorodianC04}. 

The mathematical concept \textit{topology} describes spaces up to continuous deformations. These deformations are bending, stretching and rotating, however not glueing or tearing. Two spaces are said to be equivalent if one can be transformed into the other by a continuous deformation. Therefore, topological spaces cannot be described using distances between points. The space can be described in terms of paths, loops and holes. Using simplicial homology, we can compute the number of holes of a space in each dimension. 

A simplex $S$ is a generalisation of a triangle, formed by pairwise connected points. The zero-dimensional simplices are points, one-dimensional simplices are lines, two-dimensional simplices are triangles, three-dimensional simplices are tetrahedrons and so on. Simplices themselves can be joined together to form a simplicial complex. Within a simplical complex, all faces of the complex are in the complex and the intersection of two simplices is either empty or a face of both simplices.
To compute the homology of simplicial complex, all cycles of simplices are considered and each closed cycle that does not bound another simplex is called a hole. The boundary map $\partial_k\colon \Delta_k(S)\rightarrow\Delta_{k-1}(S)$ takes a $k$-simplex to its boundary, which is a $k-1$-simplex. This map gives rise to the $k$-th homology:
\begin{equation}
\label{eq:Homology}
\mathcal{H}_k(S)=\left(\ker\partial_k\right)/\left(\im\partial_{k+1}\right).
\end{equation}

From the $k$-th homology, the $k$-th Betti-number can be calculated:
\begin{equation}
\label{eq:BettiNumber}
\beta_k=\text{rank }\mathcal{H}_k(S).
\end{equation}
Intuitively, the $k$-th Betti-number is the maximum amount of cuts that can be made before a $k$-dimensional surface is split into two parts.

An $n$-dimensional finite point cloud is a finite subset of $\mathbb{R}^n$ together with a metric $d$. Given a strictly increasing sequence of positive real numbers $s=\left\lbrace\epsilon_i\right\rbrace_{i=1}^n$, the ordered sequence $\mathcal{F}_s=\left\lbrace S_{\epsilon_i}(X)\right\rbrace _{i=1}^n$ is a filtration of simplicial complexes $S_{\epsilon_i}(X)$ on the point cloud $X$. For such a filtration $\mathcal{F}_s$, for any $p\in\mathbb{Z}_{>0}$, the $k$-th $p$-persistent homology group of the $i$-th simplicial complex is defined by 
\begin{equation}
\label{eq:PersistentHomology}
\mathcal{H}_k^{i,p}(S_{\epsilon_i})=\left(\ker \partial _k^i\right)/\left(\im\partial _{k+1}^{i+p}\cap \ker\partial_k^i\right).
\end{equation}
Here, $\partial_k^i:\Delta_k(S_{\epsilon_i}) \rightarrow\Delta_{k-1}(S_{\epsilon_i})$ denotes the boundary operator to the chain complex of the $i$-th simplicial complex. The $k$-th $p$-persistent Betti-number of the $i$-th simplicial complex is computed in the same way as the Betti-number, 
\begin{equation}
\label{eq:PersistentBettiNumber}
\beta_k^{i,p}=\text{rank }\mathcal{H}_k^{i,p}(S_{\epsilon_i}).
\end{equation}

The persistent homology is used to compute the supposed homology and gives an indication on how accurate an approximation is. It shows how holes exist over time: holes that live longest are most likely to exist in the underlying space and holes that live for a short time are likely to be noise. 

\section{\label{sec:quantumAlgorithmPH}A quantum algorithm for computing the persistent homology}
In \cite{LloydGZ16} a quantum algorithm is presented to cluster data sets based on the persistent Betti-numbers. Computing the persistent homology and consequently the persistent Betti-numbers, requires the construction of a simplicial complex on a set of $n$ data points. Computing all the Betti-numbers for all orders $k$ using the best known classical algorithms, is exponential in $n$ \cite{Basu03a}.

The algorithm in \cite{LloydGZ16} uses reduced simplicial complex. 
For these complexes, an exponential speed-up is obtained due to the efficient representation of a simplicial complex with $\mathcal{O}(2^n)$ simplices in only $\mathcal{O}(n)$ qubits. 

To construct these simplices, first the pairwise distances between all data points must be computed and then stored in a quantum state. These distances are necessary as a Vietoris-Rips complex \cite{Vietoris27} is constructed on the data points for a scale length $\epsilon$. Each simplex is one-hot encoded for the vertices in the complex and Grover's algorithm is used to construct a quantum state consisting of all $k$-simplices in the complex at scale $\epsilon$. 

Next, boundary maps $\partial_k$ are defined between spaces spanned by the vectors of the $k$-simplices at a scale~$\epsilon$. These boundary maps give rise to the $k$-th Betti-number using Equations \eqref{eq:Homology} and \eqref{eq:BettiNumber}. Typically, these Betti-numbers can be extracted by means of matrix inversion \cite{HarrowHL09}, however, this results in a quantum state. Instead, variants of this algorithm are used \cite{AbramsL99, JakschP03}, that differ from the original work in \cite{HarrowHL09} in that it only requires an approximation of the corresponding eigenvector.

\section{\label{sec:LimitationsOfAlgorithm}Limitations in finding the persistent Betti-numbers}
In the previous section the algorithm given in \cite{LloydGZ16} was briefly explained. We first show why this algorithm can in general not be used to find the persistent Betti-numbers in dimensions above the zeroth, while afterwards, we give pointers on how to use the algorithm for clustering in the zeroth dimension. 

\subsection{\label{sec:failureAboveZero}counterexample for dimensions higher than zero}
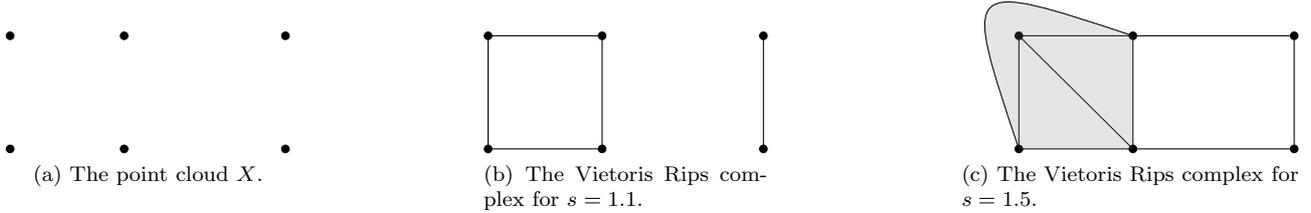
\begin{figure*}
\centering
\subfloat[The point cloud $X$.\label{fig:PointCloud}]{\begin{tikzpicture}
    \draw[fill] (0,0) circle [radius=0.05]; 
    \draw[fill] (0,1.5) circle [radius=0.05]; 
    \draw[fill] (1.5,0) circle [radius=0.05]; 
    \draw[fill] (1.5,1.5) circle [radius=0.05]; 
    \draw[fill] (1.5+2.121,0) circle [radius=0.05]; 
    \draw[fill] (1.5+2.121,1.5) circle [radius=0.05]; 
    \end{tikzpicture}}\hfill
\subfloat[The Vietoris Rips complex for $s=1.1$.\label{fig:PointCloud_1}] {\begin{tikzpicture}
    \draw[fill] (0,0) circle [radius=0.05]; 
    \draw[fill] (0,1.5) circle [radius=0.05]; 
    \draw[fill] (1.5,0) circle [radius=0.05]; 
    \draw[fill] (1.5,1.5) circle [radius=0.05]; 
    \draw[fill] (1.5+2.121,0) circle [radius=0.05]; 
    \draw[fill] (1.5+2.121,1.5) circle [radius=0.05]; 
    \draw[draw=black] (0,0) rectangle (1.5,1.5);
    \draw (1.5+2.121,0) -- (1.5+2.121,1.5);
    \end{tikzpicture}}\hfill
\subfloat[The Vietoris Rips complex for $s=1.5$.\label{fig:PointCloud_sqrt2}]{\begin{tikzpicture}
    \draw[fill] (0,0) circle [radius=0.05]; 
    \draw[fill] (0,1.5) circle [radius=0.05]; 
    \draw[fill] (1.5,0) circle [radius=0.05]; 
    \draw[fill] (1.5,1.5) circle [radius=0.05]; 
    \draw[fill] (1.5+2.121,0) circle [radius=0.05]; 
    \draw[fill] (1.5+2.121,1.5) circle [radius=0.05]; 
    \draw [draw=black] (0,0) rectangle (1.5,1.5);
    \draw (1.5,0) -- (0,1.5);
    \draw (1.5,0) -- (1.5+2.121,0) -- (1.5+2.121,1.5) -- (1.5,1.5);
    \draw[fill=gray, fill opacity=0.2] (0,0) .. controls (-.75,2.25) .. (1.5,1.5);
    \draw[draw opacity=0, fill=gray, fill opacity=0.2] (0,0) -- (1.5,0) -- (1.5,1.5) -- (0,0);
    \end{tikzpicture}}
\caption{\label{fig:1}The three simplicial complexes for scale parameters $s\in\{0, 1.1, 1.5\}$.} 
\end{figure*}

The algorithm from \cite{LloydGZ16} produces Betti-numbers $\beta_k$, which can be done efficiently on quantum computers. Afterwards, it is concluded that persistent Betti-numbers $\beta_k^{i,p}$ can be extracted from the obtained Betti-numbers $\beta_k$. We state however, that this last step is in general not possible for dimensions higher than zero. We show this by means of a counterexample. Consider the point cloud $X$ in $\mathbb{R}^2$, visualised in Fig.~\ref{fig:PointCloud}:
\begin{equation}
\label{eq:PointCloud}
\begin{split}
X = \left\{(0,0), (0,1), (1,0), (1,1),\vphantom{\sqrt{2}}\right. \\
    \left. (1+\sqrt{2},0), (1+\sqrt{2},1)\right\}.
\end{split}
\end{equation}

Furthermore, let the parameter sequence $s$ be given by $s=\{0,1.1,1.5\}$. The Vietoris-Rips complex for $s=0$ is just the point cloud given in Fig.~\ref{fig:PointCloud}, while those for $s=1.1$ and $s=1.5$ are given in Fig.~\ref{fig:PointCloud_1} and Fig.~\ref{fig:PointCloud_sqrt2}, respectively.

In the first dimension, for both scale parameters, there is precisely one hole, meaning the first Betti-number for both is $1$, i.e., $\beta_{1\mid s=1.1}=1$ and $\beta_{1\mid s=1.5} = 1$. Hence, for both scale parameters, the Betti-numbers are equal, while the values do not stem from the same homology generator. For general complexes, the Betti-numbers carry no additional information on the generators, hence from just the Betti-numbers, their persistence cannot be derived. For this example, there is no way to truthfully distinguish between the two possible value $\{0,1\}$ for $\beta_1^{1,1}$. Note that this counter-example easily extends to examples where distances between points are unique. Furthermore, the example also extends to higher dimensional topological spaces. Finally, this example for the $1$-st order Betti-number is extended to $k$-th order Betti-numbers for arbitrary $k>0$, with little extra work. 

\subsection{\label{sec:computingZerothDimension}Computing the zeroth dimensional persistent Betti-number}
While the algorithm from \cite{LloydGZ16} can in general not be used to compute the persistent Betti-numbers for dimensions higher than zero, in the zeroth dimension, it is possible. This follows as for the zeroth dimension, th persistent Betti-numbers can be computed directly from the Betti-numbers, as in the zeroth dimension, the Betti-number $\beta_0$ indicates the number of connected components, given some scale parameter. With growing scale parameter, the number of connected components does not increase. 

Let $(\epsilon_i)_i$ be a scale parameter sequence, for which the corresponding simplicial complexes are considered. In the resulting filtration of simplicial complexes for all $i\leq j$ it holds that
\begin{equation}
    \beta_0(\epsilon_i) \ge \beta_0(\epsilon_j).
    \label{eq:BetaMonotoneDecreasing}
\end{equation}

Hence, determining the persistence of Betti-numbers simplifies to determining the Betti-number for another scale parameter in the sequence. More formally, determine the $p$-persistent Betti-number of the $i$-th simplicial complex in the filtration is simply done by
\begin{equation}
    \beta_0^{i,p} = \beta_0(\epsilon_{i+p}).
    \label{eq:BetaPersistence}
\end{equation}
Clustering can consequently be done using the zeroth dimensional Betti-numbers and a bar-code on the evolution of them. An example that might prove fruitful is finding the spectrum of the graph Laplacian for clustering purposes using this algorithm, a routine that is hard classically. 

\section{\label{sec:conclusion}Summary}
In the 2016 paper by Lloyd, Garnerone and Zanardi, clustering based on topological properties of data was considered. Though not a new idea, the proposed algorithm was a quantum one, giving an exponential speed-up compared to the best known classical algorithms. 

The algorithm computes Betti-numbers in different dimensions and for different scale parameters, from which persistent Betti-numbers are deduced. We showed however, that this last step is in general not possible above the zeroth dimension. The given counterexample can easily be extended to both higher dimensional data and higher order Betti-numbers.

We also showed that the work in \cite{LloydGZ16} can still be used to compute the zeroth dimensional persistent Betti-numbers, as the persistent Betti-numbers can directly be obtained from Betti-numbers, following Eq.~\eqref{eq:BetaPersistence}. Pointers on how to use the zeroth dimensional persistent Betti-numbers are also given in this paper.

\bibliographystyle{abbrv}
\bibliography{_Limitations_of_Clustering_Using_Quantum_Persistent_Homology}

\end{document}